\documentclass{aa}

\usepackage[varg]{txfonts}

\usepackage{epsfig}
\usepackage{graphicx}

\newcommand{\beq}{\begin{equation}}
\newcommand{\eeq}{\end{equation}}
\newcommand{\beqa}{\begin{eqnarray}}
\newcommand{\eeqa}{\end{eqnarray}}
\newcommand{\beqs}{\begin{equation}\begin{split}}
\newcommand{\eeqs}{\end{split}\end{equation}}

\newcommand{\remove}[1]{}

\newcommand{\ii}{{\rm i}}

\newcommand{\vl}{{\bf l}}
\newcommand{\vn}{{\bf n}}

\newcommand{\vv}{{\bf v}}

\newcommand{\vk}{{\bf k}}
\newcommand{\vp}{{\bf p}}

\newcommand{\tdelta}{{\tilde{\delta}}}

\newcommand{\tlambda}{{\tilde{\lambda}}}

\newcommand{\tW}{{\tilde{W}}}

\newcommand{\bea}{\begin{array}}
\newcommand{\ea}{\end{array}}

\begin{document}

\title{Consistency relations for large-scale structures: Applications for the integrated Sachs-Wolfe effect and the kinematic Sunyaev-Zeldovich effect}

\titlerunning{Consistency relations for the ISW and kSZ effects}

\author{
Luca Alberto Rizzo \inst{\ref{inst1}}
\and
David F. Mota \inst{\ref{inst2}}
\and
Patrick Valageas \inst{\ref{inst1}}
}

\institute{
Institut de Physique Th\'eorique, CEA, IPhT, F-91191 Gif-sur-Yvette, C\'edex, France
\label{inst1}
\and
Institute of Theoretical Astrophysics, University of Oslo, 0315 Oslo, Norway
\label{inst2}
}

\date{\today}

\abstract{
Consistency relations of large-scale structures provide exact nonperturbative results
for cross-correlations of cosmic fields in the squeezed limit. They only depend
on the equivalence principle and the assumption of Gaussian initial conditions,
and remain nonzero at equal times for cross-correlations of density fields with velocity or
momentum fields, or with the time derivative of density fields. 
We show how to apply these relations to observational probes that involve the integrated 
Sachs-Wolfe effect or the kinematic Sunyaev-Zeldovich effect.
In the squeezed limit, this allows us to express the three-point cross-correlations, 
or bispectra, of two galaxy or matter density fields, or weak lensing convergence
fields, with the secondary Cosmic Microwave Background (CMB) distortion 
in terms of products of a linear and a nonlinear power spectrum. 
In particular, we find that cross-correlations with the integrated Sachs-Wolfe effect
show a specific angular dependence.
These results could be used to test the equivalence principle and the primordial
Gaussianity, or to check the modeling of large-scale structures.
}

\keywords{Cosmology -- large-scale structure of the Universe}

\maketitle

\section{Introduction}
\label{sec:introduction}

Measuring statistical properties of cosmological structures is not only an efficient tool 
to describe and understand the main components of our Universe, but also it is a powerful 
probe of possible new physics beyond the standard 
$\Lambda$-Cold Dark Matter ($\Lambda$CDM) concordance model. 
However, on large scales, cosmological structures are described by perturbative methods, 
while smaller scales are described by phenomenological models or studied with numerical 
simulations. It is therefore difficult to obtain accurate predictions
on the full range of scales probed by galaxy and lensing surveys.
Furthermore, if we consider galaxy density fields, theoretical predictions remain sensitive 
to the galaxy bias, which involves phenomenological modeling of star formation, 
even if we use cosmological numerical simulations. As a consequence, exact analytical 
results that go beyond low-order perturbation theory and also apply to biased tracers 
are very rare. 

Recently, some exact results have  been obtained 
\citep{Kehagias2013,Peloso2013,Creminelli2013,Kehagias2014c,Peloso2014,Creminelli2014a,Valageas2014a,Horn2014,Horn2015}
in the form of ``kinematic consistency relations''. 
They relate the $(\ell+n)$-density correlation, with $\ell$ large-scale wave numbers
and $n$ small-scale wave numbers, to the $n$-point small-scale density correlation.
These relations, obtained at the leading order over the large-scale wave numbers,
arise from the equivalence principle (EP) and the assumption of Gaussian initial conditions. 
The equivalence principle ensures that small-scale structures respond to a large-scale 
perturbation by a uniform displacement, while primordial Gaussianity provides a simple
relation between correlation and response functions (see \citet{Valageas2016} for the
additional terms associated with non-Gaussian initial conditions).
Therefore, such relations express a kinematic effect that vanishes 
for equal-times statistics, as a uniform displacement has no impact on the statistical 
properties of the density field observed at a given time.

In practice, it is, however, difficult to measure different-times density correlations 
and it would therefore be useful to obtain relations that remain nonzero at equal times.
One possibility to overcome such a problem is to go to higher orders and take into account 
tidal effects, which at leading order are given by the response of small-scale structures 
to a change in the background density. Such an approach, however, introduces some 
additional approximations \citep{Valageas2014b,Kehagias2014,Nishimichi2014}.

Fortunately, it was recently noticed that by cross-correlating density fields with velocity 
or momentum fields, or with the time derivative of the density field, one obtains
consistency relations that do not vanish at equal times \citep{Rizzo2016}. 
Indeed, the kinematic effect
modifies the amplitude of the large-scale velocity and momentum fields, while the  
time derivative of the density field is obviously sensitive to different-times effects.

In this paper, we investigate the observational applicability of these new relations.
We consider the lowest-order relations, which relate three-point cross-correlations 
or bispectra in the squeezed limit to products of a linear and a nonlinear power spectrum.
To involve the non-vanishing consistency relations, we study two observable quantities,
the secondary anisotropy $\Delta_{\rm ISW}$ of the cosmic microwave background (CMB) 
radiation due to the integrated Sachs-Wolfe effect (ISW), and the secondary anisotropy 
$\Delta_{\rm kSZ}$ due to the kinematic Sunyaev-Zeldovich (kSZ) effect.
The first process, associated with the motion of CMB photons through time-dependent
gravitational potentials, depends on the time derivative of the matter density field.
The second process, associated with the scattering of CMB photons by free electrons,
depends on the free electrons velocity field.
We investigate the cross correlations of these two secondary anisotropies with
both galaxy density fields and the cosmic weak lensing convergence.

This paper is organized as follows. In Section~\ref{sec:consistency_rel} we recall
the consistency relations of large-scale structures that apply to density,
momentum, and momentum-divergence (i.e., time derivative of the density) fields.
We describe the various observational probes that we consider in this paper
in Section~\ref{sec:observable}.
We study the ISW effect in Section~\ref{sec:consi-relation-ISW} and the
kSZ effect in Section~\ref{sec:consi-relation-kSZ}.
We conclude in Section~\ref{sec:Conclusions}.

\section{Consistency relations for large-scale structures}
\label{sec:consistency_rel}

\subsection{Consistency relations for density correlations}
\label{sec:consistency_rel-density}

As described in recent works \citep{Kehagias2013,Peloso2013,Creminelli2013,Kehagias2014c,Peloso2014,Creminelli2014a,Valageas2014a,Horn2014,Horn2015},
it is possible to obtain exact relations between density correlations of different orders
in the squeezed limit, where some of the wavenumbers are in the linear regime and far
below the other modes that may be strongly nonlinear.
These ``kinematic consistency relations'', obtained at the leading order over the large-scale 
wavenumbers, arise from the equivalence principle and the assumption of Gaussian
primordial perturbations. They express the fact that at leading order where
a large-scale perturbation corresponds to a linear gravitational potential (hence
a constant Newtonian force) over the extent of a small-size structure, the latter falls
without distortions in this large-scale potential.

Then, in the squeezed limit $k \to 0$, the correlation between one large-scale density
mode $\tilde\delta(\vk)$ and $n$ small-scale density modes $\tdelta(\vk_j)$
can be expressed in terms of the $n$-point small-scale correlation, as
\beqa
&& \hspace{-0.5cm} \langle \tdelta(\vk,\eta) \prod_{j=1}^n \tdelta(\vk_j,\eta_j) 
\rangle_{k \rightarrow 0}'  = - P_L(k,\eta)  \langle \prod_{j=1}^n \tdelta(\vk_j,\eta_j) \rangle' 
\nonumber \\
&& \times \sum_{i=1}^{n} \frac{D(\eta_i)}{D(\eta)} \frac{\vk_i \cdot \vk}{k^2} ,
\label{consistency_relation_delta}
\eeqa
where the tilde denotes the Fourier transform of the fields,
$\eta$ is the conformal time,
$D(\eta)$ is the linear growth factor, the prime in 
$\langle \dots \rangle'$ denotes that we factored out the Dirac factor,
$\langle \dots \rangle = \langle \dots \rangle' \delta_D (\sum \vk_j)$,
and  $P_L(k)$ is the linear matter power spectrum. 
It is worth stressing that these relations are valid even in the nonlinear regime and for biased 
galaxy fields $\tdelta_g(\vk_j)$.
The right-hand side gives the squeezed limit of the $(1+n)$ correlation
at the leading order, which scales as $1/k$. 
It vanishes at this order at equal times, because of the constraint associated 
with the Dirac factor $\delta_D(\sum \vk_j)$.

The geometrical factors $(\vk_i \cdot \vk)$ vanish if $\vk_i \perp \vk$.
Indeed, the large-scale mode induces a uniform displacement along the direction of $\vk$.
This has no effect on small-scale plane waves of wavenumbers $\vk_i$ with
$\vk_i \perp \vk$, as they remain identical after such a displacement.
Therefore, the terms in the right-hand side of Eq.(\ref{consistency_relation_delta})
must vanish in such orthogonal configurations, as we can check from the explicit
expression.

The simplest relation that one can obtain from Eq.\eqref{consistency_relation_delta} 
is for the bispectrum with $n=2$,
\beqa
&& \hspace{-0.5cm} \langle \tdelta(\vk,\eta) \tdelta_g(\vk_1,\eta_1) \tdelta_g(\vk_2,\eta_2) 
\rangle_{k\rightarrow 0}' =  - P_L(k,\eta) \frac{\vk_1\cdot \vk}{k^2} \nonumber \\ 
&&  \times \langle \tilde\delta_g(\vk_1,\eta_1) \tilde\delta_g(\vk_2,\eta_2) \rangle' 
\frac{D(\eta_1) - D(\eta_2)}{D(\eta)} ,
\label{bispectrum-delta-unequal} 
\eeqa
where we used that $\vk_2=-\vk_1-\vk \to - \vk_1$. 
For generality, we considered here the small-scale fields $\tdelta_g(\vk_1)$ and
$\tdelta_g(\vk_2)$ to be associated with biased tracers such as galaxies.
The tracers associated with $\vk_1$ and $\vk_2$ can be different and have different
bias.
At equal times the right-hand side of Eq.(\ref{bispectrum-delta-unequal}) vanishes,
as recalled above.

\subsection{Consistency relations for momentum correlations}
\label{sec:consistency_rel-p}

The density consistency relations (\ref{consistency_relation_delta}) express the uniform 
motion of small-scale structures by large-scale modes.
This simple kinematic effect vanishes for equal-time correlations of the density field,
precisely because there are no distortions, while there is a nonzero effect at different
times because of the motion of the small-scale structure between different times.
However, as pointed out in \citet{Rizzo2016}, it is possible to obtain
nontrivial equal-times results by considering velocity or momentum fields, which
are not only displaced but also see their amplitude affected by the large-scale mode.
Let us consider the momentum $\vp$ defined by
\beq
\vp = (1+\delta)\vv ,
\eeq
where $\vv$ is the peculiar velocity.
Then, in the squeezed limit $k \to 0$, the correlation between one large-scale density
mode $\tilde\delta(\vk)$, $n$ small-scale density modes $\tdelta(\vk_j)$, and
$m$ small-scale momentum modes $\tilde{\vp}(\vk_j)$ 
can be expressed in terms of $(n+m)$ small-scale correlations, as
\beqa
&& \hspace{-0.5cm} \langle \tdelta(\vk,\eta) \prod_{j=1}^n \tdelta(\vk_j,\eta_j) 
\prod_{j=n+1}^{n+m} \tilde{\vp}(\vk_j,\eta_j) \rangle_{k \rightarrow 0}'  = - P_L(k,\eta)  
\nonumber \\
&& \times \biggl \lbrace \langle \prod_{j=1}^n \tdelta(\vk_j,\eta_j) \prod_{j=n+1}^{n+m} 
\tilde{\vp}(\vk_j,\eta_j) \rangle' 
\sum_{i=1}^{n+m} \frac{D(\eta_i)}{D(\eta)} \frac{\vk_i \cdot \vk}{k^2} \nonumber \\
&& + \sum_{i=n+1}^{n+m} \frac{(dD/dn)(\eta_i)}{D(\eta)} 
\langle \prod_{j=1}^n 
\tdelta(\vk_j,\eta_j) \prod_{j=n+1}^{i-1} \tilde{\vp}(\vk_j,\eta_j) \nonumber \\
&& \times \left( \ii \frac{\vk}{k^2} [ \delta_D(\vk_i) + \tdelta(\vk_i,\eta_i) ] \right) 
\prod_{j=i+1}^{n+m} \tilde{\vp}(\vk_j,\eta_j) \rangle' 
  \biggl \rbrace .
\label{consistency_relation_p}
\eeqa
These relations are again valid in the nonlinear regime and for biased 
galaxy fields $\tdelta_g(\vk_j)$ and $\tilde{\vp}_g(\vk_j)$.
As for the density consistency relation (\ref{consistency_relation_delta}),
the first term vanishes at this order at equal times.
The second term, however, which arises from the $\tilde{\vp}$ fields only,
remains nonzero. This is due to the fact that $\tilde{\vp}$ involves the velocity, 
the amplitude of which is affected by the motion induced by the large-scale mode.

The simplest relation associated with Eq.\eqref{consistency_relation_p} 
is the bispectrum among two density-contrast fields and one momentum field,
\beqa
&& \hspace{-0.5cm} \langle \tdelta(\vk,\eta) \tdelta_g(\vk_1,\eta_1) \tilde{\vp}_g(\vk_2,\eta_2) 
\rangle_{k\rightarrow 0}' =  - P_L(k,\eta) \nonumber \\ 
&&  \times \bigg( \frac{\vk_1\cdot \vk}{k^2} \langle \tilde\delta_g(\vk_1,\eta_1) \tilde\vp_g(\vk_2,\eta_2) \rangle' \frac{D(\eta_1) - D(\eta_2)}{D(\eta)}    \nonumber  \\
&&  + \ii \frac{\vk}{k^2} \langle \tilde\delta_g(\vk_1,\eta_1) \tilde\delta_g(\vk_2,\eta_2) \rangle' 
\frac{1}{D(\eta)} \frac{dD}{d\eta}(\eta_2) \bigg) .
\label{bispectrum-p-unequal} 
\eeqa
For generality, we considered here the small-scale fields $\tdelta_g(\vk_1)$ and
$\tilde\vp_g(\vk_2)$ to be associated with biased tracers such as galaxies,
and the tracers associated with $\vk_1$ and $\vk_2$ can again be different and have 
different bias.
At equal times, Eq.\eqref{bispectrum-p-unequal} reads as
\beq 
\langle \tdelta(\vk) \tdelta_g(\vk_1) \tilde\vp_g(\vk_2) \rangle_{k\rightarrow 0}'  
= - \ii \frac{\vk}{k^2} \frac{d\ln D}{d\eta} P_L(k) P_g(k_1) ,
\label{bispectrum_p}
\eeq
where $P_g(k)$ is the galaxy nonlinear power spectrum and
we omitted the common time dependence. 
This result does not vanish thanks to the term generated by $\tilde\vp$
in the consistency relation (\ref{bispectrum-p-unequal}).

\subsection{Consistency relations for momentum-divergence correlations}
\label{sec:consistency_rel-lambda}

In addition to the momentum field $\vp$, we can consider its divergence
$\lambda$, defined by
\beq
\lambda \equiv \nabla \cdot \left[ (1+ \delta) \vv \right] = - \frac{\partial\delta}{\partial\eta} .
\label{lambda-def}
\eeq
The second equality expresses the continuity equation, that is, the conservation of matter.
In the squeezed limit we obtain from Eq.(\ref{consistency_relation_p})
\citep{Rizzo2016}
\beqa
&& \hspace{-0.5cm} \langle \tdelta(\vk,\eta) \prod_{j=1}^n \tdelta(\vk_j,\eta_j) 
\prod_{j=n+1}^{n+m} \tlambda(\vk_j,\eta_j) \rangle_{k \rightarrow 0}'  = - P_L(k,\eta)  
\nonumber \\
&& \times \biggl \lbrace \langle \prod_{j=1}^n \tdelta(\vk_j,\eta_j) \prod_{j=n+1}^{n+m} 
\tlambda(\vk_j,\eta_j) \rangle' 
\sum_{i=1}^{n+m} \frac{D(\eta_i)}{D(\eta)} \frac{\vk_i \cdot \vk}{k^2} \nonumber \\
&& \hspace{0.5cm} - \sum_{i=n+1}^{n+m} \langle \tdelta(\vk_i,\eta_i) \prod_{j=1}^n 
\tdelta(\vk_j,\eta_j) \prod_{\substack{j=n+1 \\ j\neq i}}^{n+m} \tlambda(\vk_j,\eta_j) \rangle' 
\nonumber \\
&& \hspace{1cm} \times \frac{(d D/d\eta)(\eta_i)}{D(\eta)} \frac{\vk_i\cdot\vk}{k^2}  \biggl \rbrace .
\label{consistency_relation_lambda}
\eeqa
These relations can actually be obtained by taking derivatives with respect to the
times $\eta_j$ of the density consistency relations (\ref{consistency_relation_delta}),
using the second equality (\ref{lambda-def}).
As for the momentum consistency relations (\ref{consistency_relation_p}),
these relations remain valid in the nonlinear regime and for biased 
small-scale fields $\tdelta_g(\vk_j)$ and $\tlambda_g(\vk_j)$.
The second term in Eq.(\ref{consistency_relation_lambda}), which arises from 
the $\tlambda$ fields only, remains nonzero at equal times.
This is due to the fact that $\lambda$ involves the velocity or the time-derivative of the density, 
which probes the evolution between (infinitesimally close) different times.

The simplest relation associated with Eq.\eqref{consistency_relation_lambda} 
is the bispectrum among two density-contrast fields and one momentum-divergence 
field,
\beqa
&& \hspace{-0.5cm} \langle \tdelta(\vk,\eta) \tdelta_g(\vk_1,\eta_1) \tlambda_g(\vk_2,\eta_2) 
\rangle_{k\rightarrow 0}' =  - P_L(k,\eta) \frac{\vk_1\cdot \vk}{k^2} \nonumber \\ 
&&  \times \bigg( \langle \tilde\delta_g(\vk_1,\eta_1) \tilde\lambda_g(\vk_2,\eta_2) \rangle' 
\frac{D(\eta_1) - D(\eta_2)}{D(\eta)}   \nonumber  \\
&&  + \langle \tilde\delta_g(\vk_1,\eta_1) \tilde\delta_g(\vk_2,\eta_2) \rangle' 
\frac{1}{D(\eta)} \frac{dD}{d\eta}(\eta_2) \bigg) .
\label{bispectrum-lambda-unequal} 
\eeqa
At equal times, Eq.\eqref{bispectrum-lambda-unequal} reads as
\beq 
\langle \tdelta(\vk) \tdelta_g(\vk_1) \tlambda_g(\vk_2) \rangle_{k\rightarrow 0}'  
= -\frac{\vk_1\cdot\vk}{k^2} \frac{d\ln D}{d\eta} P_L(k) P_g(k_1) .
\label{bispectrum_lambda}
\eeq

\section{Observable quantities}
\label{sec:observable}

To test cosmological scenarios with the consistency relations of large-scale structures
we need to relate them to observable quantities.
We describe in this section the observational probes that we consider in this paper.
We use the galaxy numbers counts or the weak lensing convergence to probe
the density field.
To apply the momentum consistency relations (\ref{bispectrum_p}) and
(\ref{bispectrum_lambda}), we use the ISW effect to probe the momentum divergence
$\lambda$ (more precisely the time derivative of the gravitational potential and matter
density) and the kSZ effect to probe the momentum $\vp$.

\subsection{Galaxy number density contrast $\delta_g$}
\label{sec:galaxy-contrast}

From galaxy surveys we can typically measure the galaxy density contrast within a
redshift bin, smoothed with a finite-size window on the sky,
\beq
\delta_g^s(\vec\theta) = \int d \vec\theta{\,'} \, W_{\Theta}(| \vec\theta{\,'} - \vec\theta |) 
\int d\eta \, I_g(\eta) \delta_g[r, r\vec\theta{\,'} ; \eta] ,
\label{deltas_g-line}
\eeq
where $W_{\Theta}(| \vec\theta{\,'} - \vec\theta |)$ is a 2D symmetric window function centered 
on the direction $\vec\theta$ on the sky, of characteristic angular radius $\Theta$,
$I_g(\eta)$ is the radial weight along the line of sight associated with a normalized
galaxy selection function $n_g(z)$,
\beq
I_g(\eta) = \left | \frac{d z}{d \eta} \right | n_g(z)  ,
\label{Ig-def}
\eeq
$r=\eta_0-\eta$ is the radial comoving coordinate along the line of sight,
and $\eta_0$ is the conformal time today.
Here and in the following we use the flat sky approximation, and $\vec\theta$ is
the 2D vector that describes the direction on the sky of a given line of sight.
The superscript ``s'' in $\delta^s_g$ denotes that we smooth the galaxy density contrast 
with the finite-size window $W_{\Theta}$.
Expanding in Fourier space, we can write the galaxy density contrast as
\beqa
\delta^s_g(\vec\theta) &= & \int d \vec\theta{\,'} \, W_{\Theta}(| \vec\theta{\,'} - \vec\theta |)
\int d\eta \, I_g(\eta) \nonumber \\
&& \times \int d\vk \; e^{\ii k_{\parallel} r + \ii \vk_{\perp} \cdot r \vec\theta{\,'}} \,
\tdelta_g(\vk,\eta), 
\eeqa
where $k_{\parallel}$ and $\vk_{\perp}$ are respectively the parallel and the perpendicular
components of the 3D wavenumber $\vk = (k_{\parallel} , \vk_{\perp})$
(with respect to the reference direction $\vec\theta=0$, and we work in the small-angle
limit $\theta \ll 1$). 
Defining the 2D Fourier transform of the window $W_{\Theta}$ as
\beq
\tilde{W}_{\Theta}(|\vec{\ell}|) = \int d \vec\theta \; 
e^{-\ii \vec{\ell} \cdot \vec\theta} W_{\Theta}(| \vec\theta |) ,
\label{tW-def}
\eeq
we obtain
\beq
\delta^s_g( \vec\theta ) = \int d\eta \, I_g(\eta) \int d\vk \,
\tW_{\Theta}(k_{\perp} r) e^{\ii k_{\parallel} r + \ii \vk_{\perp} \cdot r \vec\theta} \,
\tdelta_g(\vk,\eta) .
\label{deltas_g-tW}
\eeq

\subsection{Weak lensing convergence $\kappa$}
\label{sec:lensing}

From weak lensing surveys we can measure the weak lensing convergence, given  
in the Born approximation by
\beq
\kappa^s(\vec\theta) = \int \!\! d\vec\theta{\,'} W_{\Theta}(|\vec\theta{\,'}-\vec\theta|) 
\int \!\! d\eta \; r \, g(r) \nabla^2 \frac{\Psi+\Phi}{2}[r,r\vec\theta{\,'};\eta] ,
\label{kappa-def}
\eeq
where $\Psi$ and $\Phi$ are the Newtonian gauge gravitational potentials
and the kernel $g(r)$ that defines the radial depth of the survey is
\beq
g(r) = \int_r^{\infty} dr_s \frac{dz_s}{dr_s} n_g(z_s) \frac{r_s-r}{r_s} ,
\label{gr-lensing-def}
\eeq
where $n_g(z_s)$ is the redshift distribution of the source galaxies.
Assuming no anisotropic stress, that is, $\Phi=\Psi$, and using the Poisson equation,
\beq
\nabla^2 \Psi = 4 \pi {\cal G}_{\rm N} \bar\rho_0 \delta/a ,
\label{Poisson-x}
\eeq
where ${\cal G}_{\rm N}$ is the Newton constant, $\bar\rho_0$ is the mean matter density 
of the Universe today, and $a$ is the scale factor, we obtain
\beq
\kappa^s(\vec\theta) = \int d\eta \, I_{\kappa}(\eta) \int d\vk \,
\tW_{\Theta}(k_{\perp} r) e^{\ii k_{\parallel} r + \ii \vk_{\perp} \cdot r \vec\theta} \,
\tdelta(\vk,\eta) ,
\label{kappas-tW}
\eeq
with
\beq
I_{\kappa}(\eta) = 4 \pi {\cal G}_{\rm N} \bar\rho_0 \frac{r g(r)}{a} .
\label{Ikappa-def}
\eeq

\subsection{ISW secondary anisotropy $\Delta_{\rm ISW}$}
\label{sec:ISW}

From Eq.(\ref{lambda-def}) $\lambda$ can be obtained
from the momentum divergence or from the time derivative of the density contrast.
These quantities are not as directly measured from galaxy surveys as density contrasts.
However, we can relate the time derivative of the density contrast to the ISW effect,
which involves the time derivative of the gravitational potential.
Indeed, the secondary cosmic microwave background temperature anisotropy due to the 
integrated Sachs-Wolfe effect along the direction $\vec\theta$ reads as 
\citep{Garriga2004}
\beqa
\Delta_{\rm ISW}(\vec\theta) & = & \int d \eta \, e^{-\tau(\eta)} 
\left( \frac{\partial \Psi}{\partial \eta} + \frac{\partial \Phi}{\partial \eta} \right) 
[r,r \vec\theta;\eta] \nonumber \\
& = & 2 \int d \eta \,  e^{-\tau(\eta)} \frac{\partial \Psi}{\partial \eta} [r,r\vec\theta;\eta] ,
\label{Delta-ISW}
\eeqa
where $\tau(\eta)$ is the optical depth, which takes into account the 
possibility of late reionization, and in the second line we assumed no anisotropic stress, 
that is, $\Phi=\Psi$.
We can relate $\Delta_{\rm ISW}$ to $\lambda$ through the Poisson equation
(\ref{Poisson-x}), which reads in Fourier space as
\beq
-k^2 \tilde{\Psi} = 4 \pi {\cal G}_{\rm N} \bar{\rho}_0 \tilde{\delta} /a .
\label{Poisson-k}
\eeq
This gives
\beq
\frac{\partial\tilde\Psi}{\partial\eta} = \frac{4\pi {\cal G}_{\rm N} \bar\rho_0}{k^2 a} 
( \tlambda + {\cal H} \tdelta ) ,
\label{lambda-Psi}
\eeq
where ${\cal H} = d\ln a / d \eta$ is the conformal expansion rate.
Integrating the ISW effect $\delta_{\rm ISW}$ over some finite-size window
on the sky, we obtain, as in Eq.(\ref{deltas_g-tW}),
\beqa
\Delta^s_{\rm ISW}( \vec\theta ) & = & \int d\eta \, I_{\rm ISW}(\eta) 
\int d\vk \, \tW_{\Theta}(k_{\perp} r) e^{\ii k_{\parallel} r + \ii \vk_{\perp} \cdot r \vec\theta} 
\nonumber \\
&& \times \frac{\tlambda + {\cal H} \tdelta}{k^2} ,
\label{Deltas-ISW}
\eeqa
with
\beq
I_{\rm ISW}(\eta) = 8\pi {\cal G}_{\rm N} \bar\rho_0 \frac{e^{-\tau}}{a} .
\label{I-ISW-def}
\eeq

\subsection{Kinematic SZ secondary anisotropy $\Delta_{\rm kSZ}$}
\label{sec:kSZ}

Thomson scattering of CMB photons off moving free electrons in the hot galactic
or cluster gas generates secondary anisotropies
\citep{Sunyaev1980,Gruzinov1998a,Knox1998}.
The temperature perturbation, $\Delta_{\rm kSZ} = \delta T/T$,
due to this kinematic Sunyaev Zeldovich (kSZ) effect, is
\beq
\Delta_{\rm kSZ}(\vec\theta) = - \! \int \!\!  d \vl \cdot \vv_e \sigma_T n_e e^{-\tau} 
= \int \!\! d\eta \, I_{\rm kSZ}(\eta) \vn(\vec\theta) \cdot \vp_e ,
\label{kSZ-def}
\eeq
where $\tau$ is again the optical depth, $\sigma_T$ the Thomson cross section, 
$\vl$ the radial coordinate along the line of sight, $n_e$ the number density of free 
electrons, $\vv_e$ their peculiar velocity, and $\vn(\vec\theta)$ the radial unit vector 
pointing to the line of sight.
We also defined the kSZ kernel by
\beq
I_{\rm kSZ}(\eta) = - \sigma_T \bar{n}_e a e^{-\tau} ,
\label{I-kSZ-def}
\eeq
and the free electrons momentum $\vp_e$ as
\beq
n_e \vv_e = \bar{n}_e (1+\delta_e) \vv_e = \bar{n}_e \vp_e .
\label{pe-def}
\eeq
Because of the projection $\vn\cdot\vp_e$ along the line of sight, some care must be taken
when we smooth $\Delta_{\rm kSZ}(\vec\theta)$ over some finite-size angular window
$W_{\Theta}(|\vec\theta'-\vec\theta|)$.
Indeed, because the different lines of sight $\vec\theta'$ in the conical window are not
perfectly parallel, if we define the longitudinal and transverse momentum components
by the projection with respect to the mean line of sight $\vn(\vec\theta)$ of the circular window, 
for example, $p_{e\parallel}=\vn(\vec\theta)\cdot\vp_e$, the projection $\vn(\vec\theta') \cdot \vp_e$ 
receives contributions from both $p_{e\parallel}$ and $\vp_{e\perp}$.
In the limit of small angles we could a priori neglect the contribution associated with 
$\vp_{e\perp}$, which is multiplied by an angular factor and vanishes for a zero-size window.
However, for small but finite angles, we need to keep this contribution because fluctuations
along the lines of sight are damped by the radial integrations and vanish in the 
Limber approximation, which damps the contribution associated with $p_{e\parallel}$.

For small angles we write at linear order $\vn(\vec\theta)=(\theta_x,\theta_y,1)$, 
close to a reference direction $\vec\theta=0$. Then, the integration over the angular window
gives for the smoothed kSZ effect
\beqa
\Delta^s_{\rm kSZ}(\vec\theta) & = & \int d\eta \, I_{\rm kSZ}(\eta) \int d\vk \, 
e^{\ii \vk\cdot\vn r} \biggl [ \tilde{p}_{e\parallel} \tilde{W}_{\Theta}(k_{\perp}r)
\nonumber \\
&& - \ii \frac{\vk_{\perp}\cdot\tilde\vp_{e\perp}}{k_{\perp}} 
\tilde{W}'_{\Theta}(k_{\perp}r) \biggl ] . \;\;\;
\label{kSZ-smooth}
\eeqa
Here we expressed the result in terms of the longitudinal and transverse components of 
the wave numbers and momenta with respect to the mean line of sight $\vn(\vec\theta)$ 
of the circular window $W_{\Theta}$. 
Thus, whereas the radial unit vector is $\vn(\vec\theta)=(\theta_x,\theta_y,1)$,
we can define the transverse unit vectors as $\vn_{\perp x}=(1,0,-\theta_x)$ and
$\vn_{\perp y}=(0,1,-\theta_y)$, and we write for instance 
$\vk = k_{\perp x} \vn_{\perp x} + k_{\perp y} \vn_{\perp y} + k_{\parallel} \vn$.
We denote $\tilde{W}'_{\Theta}(\ell)=d\tilde{W}_{\Theta}/d\ell$.
The last term in Eq.(\ref{kSZ-smooth}) is due to the finite size $\Theta$ of the smoothing 
window, which makes the lines of sight within the conical beam not strictly parallel.
It vanishes for an infinitesimal window, where 
$W_{\Theta}(\vec\theta)=\delta_D(\vec\theta)$ and $\tilde{W}_{\Theta}=1$,
$\tilde{W}_{\Theta}'=0$.
We find in Section~\ref{sec:Galaxy-galaxy-kSZ} that this contribution is typically
negligible in the regime where the consistency relations apply, as the width of the small-scale
windows is much smaller than the angular size associated with the long mode.

\subsection{Comparison with some other probes}
\label{sec:Comparison}

As we explained above, in order to take advantage of the consistency relations
we use the ISW or kSZ effects because they involve the time-derivative of the
density field or the gas velocity.
The reader may then note that redshift-space distortions (RSD) also involve velocities,
but previous works that studied the galaxy density field in redshift space 
\citep{Creminelli2014a,Kehagias2014c} found that there is no equal-time effect,
as in the real-space case.
Indeed, in both real space and redshift space, the long mode only generates
a uniform change of coordinate (in the redshift-space case, this shift involves 
the radial velocity).
Then, there is no effect at equal times because such uniform shifts do not produce
distortions and observable signatures.
In contrast, in our case there is a nonzero equal-time effect because the effect
of the long mode cannot be absorbed by a simple change of coordinates.
Indeed, the kSZ effect, associated with the scattering of CMB photons by free electons
in hot ionized gas (e.g., in X-ray clusters), actually probes the velocity difference
between the rest-frame of the CMB and the hot gas. Thus, the CMB last-scattering surface
provides a reference frame and the long mode generates a velocity difference with respect
to that frame that cannot be described as a change of coordinate.
This explains why the kSZ effect makes the long-mode velocity shift observable,
without conflicting with the equivalence principle.
There is also a nonzero effect for the ISW case, because the latter involves the
time derivative of the density field, so that an equal-time statistics actually
probes different-times properties of the density field (e.g., if we write the time derivative as an infinitesimal finite difference). 

If we cross-correlate real-space and redshift-space quantities, there will also remain
a nonzero effect at equal times, because the long mode generates different shifts for
the real-space and redshift-space fields. Thus, we can consider the effect of a long mode
on small-scale correlations of the weak lensing convergence $\kappa$ with redshift-space
galaxy density contrasts $\delta^s_{g}$. However, weak lensing observables have broad
kernels along the line of sight, so that a small differential shift along the radial
direction is suppressed. In contrast, in the kSZ case the effect is directly due to the
change of velocity by the long mode, and not by the indirect impact of the change of
the radial redshift coordinate.

Another observable effect of the long mode was pointed out in \citet{Baldauf2015a}.
These authors noticed that a long mode of wave length $2\pi/k$ of the same order as the
baryon acoustic oscillation (BAO) scale, $x_{\rm BAO} \sim 110 h^{-1} {\rm Mpc}$, 
gives a different shift to galaxies separated by this distance. This produces a spread 
of the BAO peak, after we average over the long mode.
The reason why this effect is observable is that the correlation function shows
a narrow peak at the BAO scale, with a width of order 
$\Delta x_{\rm BAO} \sim 20 h^{-1} {\rm Mpc}$.
This narrow feature provides a probe of the small displacement of galaxies 
by the long mode, which would otherwise be negligible if the galaxy correlation were
a slow power law. 
As noticed above, the absence of such a narrow feature suppresses the signal associated
with cross-correlations among weak-lensing (real-space) quantities and redshift-space
quantities, because of the radial broadening of the weak-lensing probes.

This BAO probe is actually a second-order effect, in the sense of the consistency relations.
Indeed, the usual consistency relations are obtained in the large-scale limit $k \to 0$,
where the long mode generates a uniform displacement of the small-scale structures. 
In contrast, the spread of the BAO peak relies on the differential displacement between
galaxies separated by $x_{\rm BAO}$. In the Taylor expansion of the displacement
with respect to the positions of the small-scale structures, beyond the lowest-order constant 
term one takes into account the linear term over $x$, which scales as $kx.$
This is why this effect requires that $k$ be finite and not too small, of order 
$k \sim 2\pi/x_{\rm BAO}$.

\section{Consistency relation for the ISW temperature anisotropy}
\label{sec:consi-relation-ISW}

In this section we consider cross correlations with the ISW effect.
This allows us to apply the consistency relation (\ref{bispectrum-lambda-unequal}),
which involves the momentum divergence $\lambda$ and remains nonzero at equal times.

\subsection{Galaxy-galaxy-ISW correlation}
\label{sec:Galaxy-galaxy-ISW}

To take advantage of the consistency relation (\ref{bispectrum-lambda-unequal}),
we must consider three-point correlations $\xi_3$ (in configuration space) with one 
observable that involves the momentum divergence $\lambda$.
Here, using the expression (\ref{Deltas-ISW}), we study the cross-correlation between 
two galaxy density contrasts and one ISW temperature anisotropy,
\beq
\xi_3(\delta^s_g,\delta^s_{g_1},\Delta^s_{\rm ISW_2}) = \langle \delta^s_g(\vec\theta) \,
\delta^s_{g_1}(\vec\theta_1) \, \Delta^s_{\rm ISW_2}(\vec\theta_2) \rangle .
\label{correlation-delta-line}
\eeq
The subscripts $g$, $g_1$, and ${\rm ISW_2}$ denote the three lines of sight
associated with the three probes. Moreover, the subscripts $g$ and $g_1$ recall 
that the two galaxy populations associated with $\delta^s_g$ and $\delta^s_{g_1}$ 
can be different and have different bias.
As we recalled in Section~\ref{sec:consistency_rel}, the consistency relations rely
on the undistorted motion of small-scale structures by large-scale modes.
This corresponds to the squeezed limit $k \to 0$ in the Fourier-space equations 
(\ref{consistency_relation_delta}) and (\ref{consistency_relation_lambda}),
which writes more precisely as
\beq
k \ll k_L , \;\;\; k \ll k_j ,
\label{conditions-Fourier}
\eeq
where $k_L$ is the wavenumber associated with the transition between the linear and
nonlinear regimes. The first condition ensures that $\tdelta(k)$ is in the linear regime,
while the second condition ensures the hierarchy between the large-scale mode and
the small-scale modes. 
In configuration space, these conditions correspond to
\beq
\Theta \gg \Theta_L , \;\;\; \Theta \gg \Theta_j , \;\;\;
| \vec\theta - \vec\theta_j | \gg | \vec\theta_1 - \vec\theta_2 | .
\label{conditions-real-space}
\eeq
The first condition ensures that $\delta^s_g(\vec\theta)$ is in the linear regime,
whereas the next two conditions ensure the hierarchy of scales.

The expressions (\ref{deltas_g-tW}) and (\ref{Deltas-ISW}) give
\beqa
\xi_3 & \!\! = \!\! & \int d\eta d\eta_1 d\eta_2 \, I_g(\eta) I_{g_1}(\eta_1) I_{\rm ISW_2}(\eta_2) 
\nonumber \\
&& \!\! \times \!\! \int \! d\vk d\vk_1 d\vk_2 \,
\tW_{\Theta}(k_{\perp} r) \tW_{\Theta_1}(k_{1\perp} r_1) \tW_{\Theta_2}(k_{2\perp} r_2)
\nonumber \\
&& \!\! \times \; e^{\ii ( k_{\parallel} r + k_{1\parallel} r_1 + k_{2\parallel} r_2 
+ \vk_{\perp} \cdot r \vec\theta + \vk_{1\perp} \cdot r_1 \vec\theta_1
+ \vk_{2\perp} \cdot r_2 \vec\theta_2)} \nonumber \\
&& \!\! \times \langle \tdelta_g(\vk,\eta) \tdelta_{g_1}(\vk_1,\eta_1) 
\frac{\tlambda(\vk_2,\eta_2) + {\cal H}_2 \tdelta(\vk_2,\eta_2)}{k_2^2} \rangle . 
\hspace{0.8cm}
\label{ISW-consist-0}
\eeqa
The configuration-space conditions (\ref{conditions-real-space}) ensure that
we satisfy the Fourier-space conditions (\ref{conditions-Fourier}) and that we can
apply the consistency relations (\ref{bispectrum-delta-unequal}) and
(\ref{bispectrum-lambda-unequal}). This gives
\beqa
\xi_3 & = & - \int d\eta d\eta_1 d\eta_2 \, b_g(\eta) I_g(\eta) I_{g_1}(\eta_1) 
I_{\rm ISW_2}(\eta_2)  \nonumber \\
&& \hspace{-0.5cm} \times \int d\vk d\vk_1 d\vk_2 \,
\tW_{\Theta}(k_{\perp} r) \tW_{\Theta_1}(k_{1\perp} r_1) \tW_{\Theta_2}(k_{2\perp} r_2)
\nonumber \\
&& \hspace{-0.5cm} \times \; e^{\ii ( k_{\parallel} r + k_{1\parallel} r_1 + k_{2\parallel} r_2 
+ \vk_{\perp} \cdot r \vec\theta + \vk_{1\perp} \cdot r_1 \vec\theta_1
+ \vk_{2\perp} \cdot r_2 \vec\theta_2)} \nonumber \\
&& \hspace{-0.5cm} \times P_L(k,\eta) \frac{\vk_1\cdot\vk}{k^2} \delta_D(\vk+\vk_1+\vk_2) 
\nonumber \\
&& \hspace{-0.5cm} \times \Biggl( \langle \tdelta_{g_1} 
\frac{\tlambda_2 + {\cal H}_2 \tdelta_2}{k_2^2} \rangle'
\; \frac{D(\eta_1)-D(\eta_2)}{D(\eta)} \nonumber \\
&& \hspace{-0.5cm} + \langle \tdelta_{g_1} \frac{\tdelta_2}{k_2^2} 
\rangle' \; \frac{1}{D(\eta)} \frac{dD}{d\eta}(\eta_2) \Biggl) .
\label{ISW-consist-1}
\eeqa

Here we assumed that on large scales the galaxy bias is linear,
\beq
k \rightarrow 0 : \;\;\; \tilde\delta_g(\vk) = b_g(\eta) \tilde\delta(\vk) 
+ \tilde\epsilon(\vk) ,
\label{bias-def}
\eeq
where $\tilde\epsilon$ is a stochastic component that represents shot noise and the
effect of small-scale (e.g., baryonic) physics on galaxy formation.
From the decomposition (\ref{bias-def}), it is uncorrelated with the large-scale
density field \citep{Hamaus2010}, $\langle\tilde\delta(\vk)\tilde\epsilon(\vk)\rangle=0$.
Then, in Eq.(\ref{ISW-consist-1}) we neglected the term 
$\langle \tilde\epsilon \tilde\delta_{g_1} (\tilde\lambda_2+{\cal H}_2\tilde\delta_2) \rangle$.
Indeed, the small-scale local processes within the region $\vec\theta$ should be
very weakly correlated with the density fields in the distant regions $\vec\theta_1$ and
$\vec\theta_2$, which at leading order are only sensitive to the total mass within
the large-scale region $\vec\theta$. Therefore, 
$\langle \tilde\epsilon \tilde\delta_{g_1} (\tilde\lambda_2+{\cal H}_2\tilde\delta_2) \rangle$
should exhibit a fast decay at low $k$, whereas the term in Eq.(\ref{ISW-consist-1})
associated with the consistency relation only decays as $P_L(k)/k \sim k^{n_s-1}$
with $n_s \simeq 0.96$.
In Eq.(\ref{ISW-consist-1}), we also assumed that the galaxy bias $b_g$ goes 
to a constant at large scales, which is usually the case, but we could take into account
a scale dependence [by keeping the factor $b_g(k,\eta)$ in the integral over $k$].

The small-scale two-point correlations $\langle 1 \cdot 2 \rangle'$ are dominated
by contributions at almost equal times, $\eta_1 \simeq \eta_2$, as different redshifts 
would correspond to points that are separated by several Hubble radii along the lines
of sight and density correlations are negligible beyond Hubble scales.
Therefore, $\xi_3$ is dominated by the second term that does not vanish at equal times.
The integrals along the lines of sight suppress the contributions from longitudinal 
wavelengths below the Hubble radius $c/H$, while the angular windows only suppress the
wavelengths below the transverse radii $c \Theta/H$.
Then, for small angular windows, $\Theta \ll 1$, we can use Limber's approximation,
$k_{\parallel} \ll k_{\perp}$ hence $k \simeq k_{\perp}$.
Integrating over $k_{\parallel}$ through the Dirac factor 
$\delta_D(k_{\parallel}+k_{1\parallel}+k_{2\parallel})$, and next over $k_{1\parallel}$
and $k_{2\parallel}$, we obtain the Dirac factors $(2\pi)^2 \delta_D(r_1-r) \delta_D(r_2-r)$.
This allows us to integrate over $\eta_1$ and $\eta_2$ and we obtain
\beqa
\xi_3 & = & - (2\pi)^2 \int d\eta \, b_g(\eta) I_g(\eta) I_{g_1}(\eta) I_{\rm ISW_2}(\eta) 
\frac{d\ln D}{d\eta} \nonumber \\
&& \hspace{-0.5cm} \times \int d\vk_{\perp} d\vk_{1\perp} d\vk_{2\perp}
\delta_D(\vk_{\perp}+\vk_{1\perp}+\vk_{2\perp}) \tW_{\Theta}(k_{\perp} r)  \nonumber \\
&& \hspace{-0.5cm} \times \tW_{\Theta_1}(k_{1\perp} r) \tW_{\Theta_2}(k_{2\perp} r)
e^{\ii r (\vk_{\perp} \cdot \vec\theta + \vk_{1\perp} \cdot \vec\theta_1 
+ \vk_{2\perp} \cdot \vec\theta_2)} \nonumber \\
&& \hspace{-0.5cm} \times P_L(k_{\perp},\eta) 
\frac{\vk_{1\perp}\cdot\vk_{\perp}}{k_{\perp}^2 k_{2\perp}^2} P_{g_1,m}(k_{1\perp},\eta) ,
\label{ISW-consist-2}
\eeqa
where $P_{g_1,m}$ is the galaxy-matter power spectrum.
The integration over $\vk_{2\perp}$ gives
\beqa
\xi_3 & \!\! = \!\! & - (2\pi)^2 \int \!\! d\eta \, b_g I_g I_{g_1} I_{\rm ISW_2} \frac{d\ln D}{d\eta} 
\int \!\! d\vk_{\perp} d\vk_{1\perp} \tW_{\Theta}(k_{\perp} r)  \nonumber \\
&& \hspace{-0.5cm} \times \tW_{\Theta_1}(k_{1\perp} r) \tW_{\Theta_2}(k_{1\perp} r) 
P_L(k_{\perp},\eta) P_{g_1,m}(k_{1\perp},\eta) \nonumber \\
&& \hspace{-0.5cm} \times e^{\ii r [ \vk_{\perp} \cdot (\vec\theta-\vec\theta_2) 
+ \vk_{1\perp} \cdot (\vec\theta_1-\vec\theta_2)]} 
\frac{\vk_{1\perp}\cdot\vk_{\perp}}{k_{1\perp}^2 k_{\perp}^2} ,
\eeqa
and the integration over the angles of $\vk_{\perp}$ and $\vk_{1\perp}$ gives
\beqa
\xi_3 & = & \frac{(\vec\theta-\vec\theta_2) \cdot (\vec\theta_1 - \vec\theta_2)}
{| \vec\theta -\vec\theta_2 |  | \vec\theta_1 - \vec\theta_2 |} (2\pi)^4
\int d\eta \, b_g I_g I_{g_1} I_{\rm ISW_2} \frac{d\ln D}{d\eta} \nonumber \\
&& \hspace{-0.5cm} \times \int_0^{\infty} d k_{\perp} d k_{1\perp} \, \tW_{\Theta}(k_{\perp} r) 
\tW_{\Theta_1}(k_{1\perp} r) \tW_{\Theta_2}(k_{1\perp} r) \nonumber \\
&& \hspace{-0.5cm} \times P_L(k_{\perp},\eta) P_{g_1,m}(k_{1\perp},\eta)
J_1(k_{\perp} r | \vec\theta-\vec\theta_2 |) \nonumber \\ 
&& \hspace{-0.5cm} \times J_1(k_{1\perp} r | \vec\theta_1-\vec\theta_2 |) ,
\label{xi3-J1}
\eeqa
where $J_1$ is the first-order Bessel function of the first kind. 

As the expression (\ref{xi3-J1}) arises from the kinematic consistency relations, 
it expresses the response of the small-scale two-point correlation
$\langle \delta^s_{g_1}(\vec\theta_1) \, \Delta^s_{\rm ISW_2}(\vec\theta_2) \rangle$ 
to a change of the initial condition associated with the large-scale mode
$\delta^s_g(\vec\theta)$.
The kinematic effect given at the leading order by Eq.(\ref{xi3-J1}) is due to the
uniform motion of the small-scale structures by the large-scale mode.
This explains why the result (\ref{xi3-J1}) vanishes in the two following cases:

\begin{enumerate}

\item $(\vec\theta-\vec\theta_2) \perp (\vec\theta_1-\vec\theta_2)$. 
There is a nonzero response of $\langle \delta_1 \lambda_2\rangle$
if there is a linear dependence on $\delta(\vec\theta)$ of $\langle \delta_1 \lambda_2\rangle$,
so that its first derivative is nonzero. A positive (negative) $\delta(\vec\theta)$ leads 
to a uniform motion at $\vec\theta_2$ towards (away from) $\vec\theta$, along the direction
$(\vec\theta-\vec\theta_2)$.
From the point of view of $\vec\theta_1$ and $\vec\theta_2$, there is a reflection symmetry 
with respect to the axis $(\vec\theta_1-\vec\theta_2)$. For instance, if $\delta_1>0$ the density 
contrast at a position $\vec\theta_3$ typically decreases in the mean with the radius 
$|\vec\theta_3-\vec\theta_1|$, and for $\Delta \vec\theta_2 \perp (\vec\theta_1-\vec\theta_2)$ 
the points $\vec\theta_3^{\pm}= \vec\theta_2\pm\Delta\vec\theta_2$ are at the same distance 
from $\vec\theta_1$ and have the same density contrast $\delta_3$ in the mean, with typically 
$\delta_3<\delta_2$ as $|\vec\theta_3^{\pm}-\vec\theta_1| > |\vec\theta_2-\vec\theta_1|$.
Therefore, the large-scale flow along $(\vec\theta-\vec\theta_2)$ leads to a positive 
$\lambda_2=-\Delta\delta_2/\Delta\eta_2$ independently of whether the matter moves
towards or away from $\vec\theta$ (here we took a finite deviation $\Delta \vec\theta_2$).
This means that the dependence of $\langle \delta_1 \lambda_2\rangle$ on $\delta(\vec\theta)$ 
is quadratic (it does not depend on the sign of $\delta(\vec\theta)$) and the first-order response 
function vanishes.
Then, the leading-order contribution to $\xi_3$ vanishes.
(For infinitesimal deviation $\Delta \vec\theta_2$ we have 
$\lambda_2=-\partial\delta_2/\partial\eta_2=0$; by this symmetry, in the mean $\delta_2$ 
is an extremum of the density contrast along the orthogonal direction to 
$(\vec\theta_1-\vec\theta_2)$).

\item $\vec\theta_1 = \vec\theta_2$. This is a particular case of the previous configuration. 
Again, by symmetry from the viewpoint of $\delta_1$, the two points 
$\delta(\vec\theta_2+\Delta \vec\theta_2)$ and $\delta(\vec\theta_2-\Delta \vec\theta_2)$ 
are equivalent and the mean response associated with the kinematic effect vanishes.

\end{enumerate}

This also explains why Eq.(\ref{xi3-J1}) changes sign with $(\vec\theta_1-\vec\theta_2)$ 
and $(\vec\theta-\vec\theta_2)$.
Let us consider for simplicity the case where the three points are aligned and
$\delta(\vec\theta) > 0$, so that the large-scale flow points towards $\vec\theta$.
We also take $\delta_1>0$, so that in the mean the density is peaked at $\vec\theta_1$ and
decreases outwards. 
Let us take $\vec\theta_2$ close to $\vec\theta_1$, on the decreasing radial slope,
and on the other side of $\vec\theta_1$ than $\vec\theta$.
Then, the large-scale flow moves matter at $\vec\theta_2$ towards $\vec\theta_1$, so that
the density at $\vec\theta_2$ at a slightly later time comes from more outward regions 
(with respect to the peak at $\vec\theta_1$) with a lower density. This means that 
$\lambda_2 = - \partial\delta_2/\partial\eta_2$ is positive so that $\xi_3>0$.
This agrees with Eq.(\ref{xi3-J1}), as 
$(\vec\theta-\vec\theta_2) \cdot (\vec\theta_1 - \vec\theta_2)>0$ in this
geometry, and we assume the integrals over wavenumbers are dominated by the
peaks of $J_1>0$.
If we flip $\vec\theta_2$ to the other side of $\vec\theta_1$, we find on the contrary that the 
large-scale flow brings higher-density regions to $\vec\theta_2$, so that we have the change 
of signs $\lambda_2<0$ and $\xi_3<0$.
The same arguments explain the change of sign with $(\vec\theta-\vec\theta_2)$.
In fact, it is the relative direction between $(\vec\theta-\vec\theta_2)$ and 
$(\vec\theta_1-\vec\theta_2)$ that matters, measured by the scalar product 
$(\vec\theta-\vec\theta_2) \cdot (\vec\theta_1 - \vec\theta_2)$.
This geometrical dependence of the leading-order contribution to $\xi_3$ could provide
a simple test of the consistency relation, without even computing the explicit
expression in the right-hand side of Eq.(\ref{xi3-J1}).

\subsection{Three-point correlation in terms of a two-point correlation}
\label{sec:3pt-2pt}

The three-point correlation $\xi_3$ in Eq.(\ref{xi3-J1}) cannot be written as a product
of two-point correlations because there is only one integral along the line of sight that is left.
However, if the linear power spectrum $P_L(k,z)$ is already known, we may write
$\xi_3$ in terms of some two-point correlation $\xi_2$.
For instance, the small-scale cross-correlation between one galaxy density contrast and one
weak lensing convergence,
\beq
\xi_2(\delta^s_{g_1},\kappa^s_2) = \langle \delta^s_{g_1}(\vec\theta_1) 
\kappa^s_2(\vec\theta_2) \rangle 
\label{xi2-deltag-kappa-def}
\eeq
reads as
\beqa
\xi_2 & = & (2\pi)^2 \int d\eta \, I_{g_1} I_{\kappa_2} \int_0^{\infty} dk_{1\perp} k_{1\perp} \,
\tilde{F}_{\Theta_1}(k_{1\perp} r) \nonumber \\
&& \times \tilde{F}_{\Theta_2}(k_{1\perp} r) J_0(k_{1\perp} r |\vec\theta_1-\vec\theta_2|) 
P_{g_1,m}(k_{1\perp}) , \hspace{0.5cm}
\label{xi2-delta2-kappa}
\eeqa
where we again used Limber's approximation.
Here we denoted the angular smoothing windows by $\tilde{F}$ to distinguish $\xi_2$
from $\xi_3$.
Then, we can write
\beq
\xi_3 = \frac{(\vec\theta-\vec\theta_2) \cdot (\vec\theta_1 - \vec\theta_2)}
{| \vec\theta -\vec\theta_2 |  | \vec\theta_1 - \vec\theta_2 |} \xi_2 ,
\label{xi3-xi2}
\eeq
if the angular windows of the two-point correlation are chosen such that
\beqa
&& \tilde{F}_{\Theta_1}(k_{1\perp}r) \tilde{F}_{\Theta_2}(k_{1\perp}) = (2\pi)^2
\frac{I_g I_{\rm ISW_2}}{I_{\kappa_2}} b_g \frac{d\ln D}{d\eta} \nonumber \\
&& \times \left( \int_0^{\infty} d k_{\perp} \tW_{\Theta}(k_{\perp} r) 
J_1(k_{\perp} r | \vec\theta-\vec\theta_2 |) P_L(k_{\perp},\eta) \right) \nonumber \\
&& \times \frac{\tW_{\Theta_1}(k_{1\perp} r) \tW_{\Theta_2}(k_{1\perp} r) 
J_1(k_{1\perp} r | \vec\theta_1-\vec\theta_2 |)}{k_{1\perp} 
J_0(k_{1\perp} r |\vec\theta_1-\vec\theta_2|)} .
\label{F1F2-W1W2}
\eeqa
This implies that the angular windows $\tilde{F}_{\Theta_1}$ and $\tilde{F}_{\Theta_2}$
of the two-point correlation $\xi_2$ have an explicit redshift dependence.

In practice, the expression (\ref{F1F2-W1W2}) may not be very convenient.
Then, to use the consistency relation (\ref{xi3-J1}) it may be more practical
to first measure the power spectra $P_L$ and $P_{g_1,m}$ independently,
at the redshifts needed for the integral along the line of sight (\ref{xi3-J1}),
and next compare the measure of $\xi_3$ with the expression (\ref{xi3-J1})
computed with these power spectra.

\subsection{Lensing-lensing-ISW correlation}
\label{sec:Lensing-lensing-ISW}

From Eq.(\ref{xi3-J1}) we can directly obtain the lensing-lensing-ISW  three-point
correlation,
\beq
\xi_3(\kappa^s,\kappa^s_1,\Delta^s_{\rm ISW_2}) = \langle \kappa^s(\vec\theta) \,
\kappa^s_1(\vec\theta_1) \, \Delta^s_{\rm ISW_2}(\vec\theta_2) \rangle ,
\label{xi3-lens-lens-ISW}
\eeq
by replacing the galaxy kernels $b_g I_g$ and $I_{g_1}$ by the lensing convergence 
kernels $I_{\kappa}$ and $I_{\kappa_1}$,
\beqa
\xi_3 & = & \frac{(\vec\theta-\vec\theta_2) \cdot (\vec\theta_1 - \vec\theta_2)}
{| \vec\theta -\vec\theta_2 |  | \vec\theta_1 - \vec\theta_2 |} (2\pi)^4
\int d\eta \, I_{\kappa} I_{\kappa_1} I_{\rm ISW_2} \frac{d\ln D}{d\eta} \nonumber \\
&& \hspace{-0.5cm} \times \int_0^{\infty} d k_{\perp} d k_{1\perp} \, \tW_{\Theta}(k_{\perp} r) 
\tW_{\Theta_1}(k_{1\perp} r) \tW_{\Theta_2}(k_{1\perp} r) \nonumber \\
&& \hspace{-0.5cm} \times P_L(k_{\perp},\eta) P(k_{1\perp},\eta)
J_1(k_{\perp} r | \vec\theta-\vec\theta_2 |) \nonumber \\ 
&& \hspace{-0.5cm} \times J_1(k_{1\perp} r | \vec\theta_1-\vec\theta_2 |) .
\label{xi3-lens-lens-ISW-J1}
\eeqa
As compared with Eq.(\ref{xi3-J1}), the advantage of the cross-correlation with the weak 
lensing convergence $\kappa$ is that Eq.(\ref{xi3-lens-lens-ISW-J1}) involves the matter 
power spectrum $P(k_{1\perp})$ instead of the more complicated galaxy-matter 
cross power spectrum $P_{g_1,m}(k_{1\perp})$.

\subsection{Vanishing contribution to the galaxy-ISW-ISW correlation}
\label{sec:Galaxy-ISW-ISW}

In the previous Section~\ref{sec:Galaxy-galaxy-ISW}, we considered the three-point 
galaxy-galaxy-ISW correlation (\ref{correlation-delta-line}), to take advantage of the momentum 
dependence of the ISW effect (or more precisely its dependence on the time derivative 
of the density field), which gives rise to consistency relations that do not vanish at equal times.
The reader may wonder whether we could also use the galaxy-ISW-ISW correlation for the
same purpose.
From Eq.(\ref{lambda-Psi}), this three-point correlation involves 
$\langle \tilde\delta (\tilde\lambda_1+\tilde\delta_1) (\tilde\lambda_2+\tilde\delta_2) \rangle'$,
instead of $\langle \tilde\delta \tilde\delta_1 (\tilde\lambda_2+\tilde\delta_2) \rangle'$
in Eq.(\ref{ISW-consist-0}), where we use compact notations.
Thus, we obtain the combination
\beq
\langle \delta \Delta_{\rm ISW_1} \Delta_{\rm ISW_2} \rangle \propto
\langle \tilde\delta \tilde\lambda_1 \tilde\lambda_2 \rangle'
+ {\cal H} \left[ \langle \tilde\delta \tilde\lambda_1 \tilde\delta_2 \rangle' 
+ \langle \tilde\delta \tilde\delta_1 \tilde\lambda_2 \rangle' \right]
+ {\cal H}^2 \langle \tilde\delta \tilde\delta_1 \tilde\delta_2 \rangle' .
\label{galaxy-ISW-ISW}
\eeq
On the other hand, at equal times the consistency relation (\ref{consistency_relation_lambda})
writes as
\beqa
&& \hspace{-0.5cm} \langle \tdelta(\vk) \prod_{j=1}^n \tdelta(\vk_j) 
\prod_{j=n+1}^{n+m} \tlambda(\vk_j) \rangle_{k \rightarrow 0}'  = P_L(k)  \frac{D'}{D}
\sum_{i=n+1}^{n+m} \frac{\vk\cdot\vk_i}{k^2} \nonumber \\
&& \times \; \langle \tdelta(\vk_i) \prod_{j=1}^n 
\tdelta(\vk_j) \prod_{\substack{j=n+1 \\ j\neq i}}^{n+m} \tlambda(\vk_j) \rangle' ,
\label{consistency_relation_lambda_equal-times}
\eeqa
where we only keep the contributions of order $1/k$ and the second line in 
Eq.(\ref{consistency_relation_lambda}) cancels out.
The first contribution to the three-point correlation (\ref{galaxy-ISW-ISW}) reads as
\beqa
\langle \tilde\delta \tilde\lambda_1 \tilde\lambda_2 \rangle'  & = & P_L(k) \frac{D'}{D} \left[
\frac{\vk\cdot\vk_1}{k^2} \langle \tilde\delta_1 \tilde\lambda_2  \rangle' + 
\frac{\vk\cdot\vk_2}{k^2} \langle \tilde\delta_2 \tilde\lambda_1 \rangle' \right]  
\nonumber \\
& = & P_L(k) \frac{D'}{D} \frac{\vk\cdot\vk_1}{k^2} \left[ \langle \tilde\delta(\vk_1) 
\tilde\lambda(-\vk_1)  \rangle' - \langle \tilde\delta(-\vk_1) \tilde\lambda(\vk_1) \rangle' \right]  
\nonumber \\
& = & 0 .
\eeqa
Here again, we only consider the leading contribution of order $1/k$ and we use
$\vk_2=-\vk_1$ in the limit $k\to 0$. The term in the bracket in the second line vanishes
because the cross-power spectrum 
$\langle \tilde\delta(\vk) \tilde\lambda(-\vk)\rangle'  = P_{\delta,\lambda}(k)$
only depends on $|\vk|$, because of statistical isotropy. 
The second contribution to Eq.(\ref{galaxy-ISW-ISW}) reads as
\beqa
\langle \tilde\delta \tilde\lambda_1 \tilde\delta_2 \rangle' 
+ \langle \tilde\delta \tilde\delta_1 \tilde\lambda_2 \rangle' 
& = & P_L(k) \frac{D'}{D} \left[
\frac{\vk\cdot\vk_1}{k^2} \langle \tilde\delta_1 \tilde\delta_2  \rangle' + 
\frac{\vk\cdot\vk_2}{k^2} \langle \tilde\delta_2 \tilde\delta_1 \rangle' \right]  
\nonumber \\
& = & 0 .
\eeqa
The third contribution $\langle \tilde\delta \tilde\delta_1 \tilde\delta_2 \rangle'$ vanishes as usual
at equal times, as it only involves the density field.
Thus, we find that the leading-order contribution to the galaxy-ISW-ISW three-point 
correlation vanishes, in contrast with the galaxy-galaxy-ISW three-point 
correlation studied in section~\ref{sec:Galaxy-galaxy-ISW}.
This is why we focus on the three-point correlations (\ref{correlation-delta-line}) and
(\ref{xi3-lens-lens-ISW}), with only one ISW field.

This cancellation can be understood from symmetry. Let us consider the maximal case
where the points $\{\vec\theta,\vec\theta_1,\vec\theta_2\}$ are aligned. 
There is a nonzero consistency relation if the dependence of 
$\langle \lambda_1 \lambda_2\rangle'$ to $\delta(\theta)$ contains a linear term.
In the long-mode limit, this means that $\langle \lambda_1 \lambda_2\rangle'$
changes sign with the sign of the large-scale velocity flow. However, by symmetry  
$\langle \lambda_1 \lambda_2\rangle'$ does not select a left or right direction along the line 
$(\vec\theta_1,\vec\theta_2)$, so that it cannot depend on the sign of the large-scale velocity 
flow, nor on the sign of $\delta(\vec\theta)$.
In contrast, in the case of the three-point correlation (\ref{correlation-delta-line}), with only
one ISW observable, the consistency relation relies on the dependence of
$\langle \delta_1 \lambda_2 \rangle'$ on the large-scale mode $\delta$
(see the discussion after Eq.(\ref{xi3-J1})). Then, it is clear that the nonsymmetrical
quantity $\langle \delta_1 \lambda_2 \rangle'$ defines a direction along the axis
$(\vec\theta_1,\vec\theta_2)$, and a linear dependence on $\delta(\vec\theta)$
and on the sign of the large-scale velocity is expected.

\section{Consistency relation for the kSZ effect}
\label{sec:consi-relation-kSZ}

In this section we consider cross correlations with the kSZ effect.
This allows us to apply the consistency relation (\ref{bispectrum-p-unequal}),
which involves the momentum $\vp$ and remains nonzero at equal times.

\subsection{Galaxy-galaxy-kSZ correlation}
\label{sec:Galaxy-galaxy-kSZ}

In a fashion similar to the galaxy-galaxy-ISW correlation studied in
Section~\ref{sec:Galaxy-galaxy-ISW}, we consider the three-point correlation between
two galaxy density contrasts and one kSZ CMB anisotropy,
\beq
\xi_3(\delta^s_g,\delta^s_{g_1},\Delta^s_{\rm kSZ_2}) = \langle \delta^s_g(\vec\theta) \,
\delta^s_{g_1}(\vec\theta_1) \, \Delta^s_{\rm kSZ_2}(\vec\theta_2) \rangle ,
\label{correlation-delta-line-kSZ}
\eeq
in the squeezed limit given by the conditions (\ref{conditions-Fourier}) in Fourier space
and (\ref{conditions-real-space}) in configuration space.
The expressions (\ref{deltas_g-tW}) and (\ref{kSZ-smooth}) give
\beq
\xi_3 = \xi_{3\parallel} + \xi_{3\perp} ,
\eeq
with
\beqa
\xi_{3\parallel} & \!\! = \!\! & \int d\eta d\eta_1 d\eta_2 \, I_g(\eta) I_{g_1}(\eta_1) 
I_{\rm kSZ_2}(\eta_2) \int \! d\vk d\vk_1 d\vk_2 \nonumber \\
&& \hspace{-0.5cm} \times e^{\ii ( \vk \cdot \vn r + \vk_1 \cdot \vn_1 r_1 
+ \vk_2 \cdot \vn_2 r_2)} \tW_{\Theta}(k_{\perp}^{(\vn)} r) 
\tW_{\Theta_1}(k_{1\perp}^{(\vn_1)} r_1) \nonumber \\
&& \hspace{-0.5cm} \times \tW_{\Theta_2}(k_{2\perp}^{(\vn_2)} r_2)  
\langle \tdelta_g(\vk,\eta) \tdelta_{g_1}(\vk_1,\eta_1) 
\tilde{p}^{(\vn_2)}_{e\parallel})(\vk_2,\eta_2) \rangle  \hspace{0.8cm}
\label{kSZ-consistency-parallel-0}
\eeqa
and
\beqa
\xi_{3\perp} & \!\! = \!\! & -\ii \int d\eta d\eta_1 d\eta_2 \, I_g(\eta) I_{g_1}(\eta_1) 
I_{\rm kSZ_2}(\eta_2) \int \! d\vk d\vk_1 d\vk_2 \nonumber \\
&& \hspace{-0.5cm} \times e^{\ii ( \vk \cdot \vn r + \vk_1 \cdot \vn_1 r_1 
+ \vk_2 \cdot \vn_2 r_2)} \tW_{\Theta}(k_{\perp}^{(\vn)} r) 
\tW_{\Theta_1}(k_{1\perp}^{(\vn_1)} r_1) \nonumber \\
&& \hspace{-0.5cm} \times \tW_{\Theta_2}'(k_{2\perp}^{(\vn_2)} r_2)  
\langle \tdelta_g(\vk,\eta) \tdelta_{g_1}(\vk_1,\eta_1) 
\frac{\vk_{2\perp}^{(\vn_2)}\cdot\tilde{\vp}^{(\vn_2)}_{e\perp}}{k_{2\perp}^{(\vn_2)}}
(\vk_2,\eta_2) \rangle , \nonumber \\
&& 
\label{kSZ-consistency-transverse-0}
\eeqa
where we split the longitudinal and transverse contributions to Eq.(\ref{kSZ-smooth}).
Here $\{\vn,\vn_1,\vn_2\}$ are the radial unit vectors that point to the centers 
$\{\vec\theta,\vec\theta_1,\vec\theta_2\}$ of the three circular windows,
and $\{(k_{\parallel}^{(\vn)},\vk_{\perp}^{(\vn)}),(k_{1\parallel}^{(\vn_1)},\vk_{1\perp}^{(\vn_1)}),(k_{2\parallel}^{(\vn_2)},\vk_{2\perp}^{(\vn_2)})\}$ 
are the longitudinal and transverse wave numbers with respect to the associated central lines
of sight [e.g., $k_{\parallel}^{(\vn)} = \vn\cdot\vk$].

The computation of the transverse contribution (\ref{kSZ-consistency-transverse-0})
is similar to the computation of the ISW three-point correlation (\ref{ISW-consist-1}),
using again Limber's approximation. At lowest order we obtain
\beqa
\xi_{3\perp} & = & \frac{(\vec\theta-\vec\theta_1) \cdot (\vec\theta_2 - \vec\theta_1)}
{| \vec\theta -\vec\theta_1 |  | \vec\theta_2 - \vec\theta_1 |} (2\pi)^4
\int d\eta \, b_g I_g I_{g_1} I_{\rm kSZ_2} \frac{d\ln D}{d\eta} \nonumber \\
&& \hspace{-0.5cm} \times \int_0^{\infty} d k_{\perp} d k_{2\perp} \, k_{2\perp}
\tW_{\Theta}(k_{\perp} r) \tW_{\Theta_1}(k_{2\perp} r) \tW_{\Theta_2}'(k_{2\perp} r) 
\nonumber \\
&& \hspace{-0.5cm} \times P_L(k_{\perp},\eta) P_{g_1,e}(k_{2\perp},\eta)
J_1(k_{\perp} r | \vec\theta-\vec\theta_1 |) \nonumber \\ 
&& \hspace{-0.5cm} \times J_1(k_{2\perp} r | \vec\theta_2-\vec\theta_1 |) ,
\label{kSZ-transverse-1}
\eeqa
where $P_{g_1,e}$ is the galaxy-free electrons cross power spectrum.

The computation of the longitudinal contribution (\ref{kSZ-consistency-parallel-0})
requires slightly more care.
Applying the consistency relation (\ref{bispectrum-p-unequal}) gives
\beqa
\xi_{3\parallel} & = & - \int d\eta d\eta_1 d\eta_2 \, b_g(\eta) I_g(\eta) I_{g_1}(\eta_1) 
I_{\rm kSZ_2}(\eta_2) \nonumber \\
&& \hspace{-0.5cm} \times \int d\vk d\vk_1 d\vk_2 \, \tW_{\Theta}(k_{\perp}^{(\vn)} r) 
\tW_{\Theta_1}(k_{1\perp}^{(\vn_1)} r_1) \tW_{\Theta_2}(k_{2\perp}^{(\vn_2)} r_2)  
\nonumber \\
&& \hspace{-0.5cm} \times \; e^{\ii ( \vk \cdot \vn r + \vk_1 \cdot \vn_1 r_1
+ \vk_2 \cdot \vn_2 r_2)} D(\eta) P_{L0}(k)  \frac{dD}{d\eta}(\eta_2) \nonumber \\
&& \hspace{-0.5cm} \times \; \ii \frac{\vn_2\cdot\vk}{k^2} 
\langle \tdelta_{g_1} \tdelta_{e_2} \rangle' \; \delta_D(\vk+\vk_1+\vk_2) , 
\label{kSZ-consistency-1}
\eeqa
where we only kept the contribution that does not vanish at equal times, as it dominates
the integrals along the lines of sight, and we used $P_L(k,\eta) = D(\eta)^2 P_{L0}(k)$.
If we approximate the three lines of sight as parallel, we can write 
$\vn_2\cdot\vk=k_{\parallel}$, where the longitudinal and transverse directions coincide
for the three lines of sight. Then, Limber's approximation, which corresponds to the limit
where the radial integrations have a constant weight on the infinite real axis,
gives a Dirac term $\delta_D(k_{\parallel})$ and $\xi_{3\parallel}=0$ (more precisely,
as we recalled above Eq.(\ref{ISW-consist-2}), the radial integration gives
$k_{\parallel} \lesssim H/c$ while the angular window gives 
$k_{\perp} \lesssim H/(c\Theta)$ so that $k_{\parallel} \ll k_{\perp}$).
Taking into account the small angles between the different lines of sight, as for the
derivation of Eq.(\ref{kSZ-smooth}), the integration over $\vk_2$ through the Dirac factor
gives at leading order in the angles
\beqa
\xi_{3\parallel} & \!\! = \!\! & - \!\! \int \!\! d\eta d\eta_1 d\eta_2 \, b_g(\eta) 
I_g(\eta) D(\eta) I_{g_1}(\eta_1) I_{\rm kSZ_2}(\eta_2) \frac{dD}{d\eta}(\eta_2) \nonumber \\
&& \hspace{-0.5cm} \times \int \!\! dk_{\parallel} d\vk_{\perp} dk_{1\parallel} d\vk_{1\perp} 
\, \tW_{\Theta}(k_{\perp} r) \tW_{\Theta_1}(k_{1\perp} r_1) \tW_{\Theta_2}(k_{1\perp} r_2)  
\nonumber \\
&& \hspace{-0.5cm} \times \; e^{\ii [ k_{\parallel} (r-r_2) 
+ \vk_{\perp} \cdot (\vec\theta-\vec\theta_2) r_2 + k_{1\parallel} (r_1-r_2) 
+ \vk_{1\perp}\cdot (\vec\theta_1-\vec\theta_2) r_2 ]}
\nonumber \\
&& \hspace{-0.5cm} \times \; P_{L0}(k_{\perp}) P_{g_1,e}(k_{1\perp};\eta_1,\eta_2)
\ii \frac{k_{\parallel}+\vk_{\perp} \cdot (\vec\theta_2-\vec\theta)}{k_{\perp}^2} .
\label{kSZ-consistency-2}
\eeqa
We used Limber's approximation to write for instance 
$P_{L0}(k) \simeq P_{L0}(k_{\perp})$, but we kept the factor $k_{\parallel}$
in the last term, as the transverse factor $\vk_{\perp} \cdot (\vec\theta_2-\vec\theta)$,
due to the small angle between the lines of sight $\vn$ and $\vn_2$, 
is suppressed by the small angle $|\vec\theta_2-\vec\theta|$.
We again split $\xi_{3\parallel}$ over two contributions, 
$\xi_{3\parallel} = \xi_{3\parallel}^{\parallel} + \xi_{3\parallel}^{\perp}$,
associated with the factors $k_{\parallel}$ and 
$\vk_{\perp} \cdot (\vec\theta_2-\vec\theta)$ of the last term.
Let us first consider the contribution $\xi_{3\parallel}^{\parallel}$.
Writing $\ii k_{\parallel} e^{\ii k_{\parallel} (r-r_2)} = \frac{\partial}{\partial r} 
e^{\ii k_{\parallel} (r-r_2)}$, we integrate by parts over $\eta$. 
For simplicity we assume that the galaxy selection function $I_g$ vanishes
at $z=0$,
\beq
I_g(\eta_0) = 0 ,
\eeq
so that the boundary term at $z=0$ vanishes.
Then, the integrations over $k_{\parallel}$ and $k_{1\parallel}$ give a factor 
$(2\pi)^2 \delta_D(r-r_2) \delta_D(r_1-r_2)$,
and we can integrate over $\eta$ and $\eta_1$.
Finally, the integration over the angles of the transverse wavenumbers yields
\beqa
\xi_{3\parallel}^{\parallel} & = & - (2\pi)^4 \int d\eta \, \frac{d}{d\eta} 
\left[ b_g I_g D \right] I_{g_1} I_{\rm kSZ_2} \frac{dD}{d\eta} \nonumber \\
&& \hspace{-0.5cm} \times \int_0^{\infty} d k_{\perp} d k_{1\perp} \, 
\tW_{\Theta}(k_{\perp} r) \tW_{\Theta_1}(k_{1\perp} r) \tW_{\Theta_2}(k_{1\perp} r) 
\nonumber \\
&& \hspace{-0.5cm} \times \frac{k_{1\perp}}{k_{\perp}} P_{L0}(k_{\perp}) 
P_{g_1,e}(k_{1\perp},\eta) J_0(k_{\perp} r | \vec\theta-\vec\theta_2 |) \nonumber \\ 
&& \hspace{-0.5cm} \times J_0(k_{1\perp} r | \vec\theta_1-\vec\theta_2 |) ,
\label{xi3-kSZ-parallel}
\eeqa
where $J_0$ is the zeroth-order Bessel function of the first kind.
For the transverse contribution $\xi_{3\parallel}^{\perp}$ we can proceed in the same 
fashion, without integration by parts over $\eta$. This gives
\beqa
\xi_{3\parallel}^{\perp} & = & - (2\pi)^4 \int d\eta \, b_g I_g I_{g_1} I_{\rm kSZ_2} 
D \frac{dD}{d\eta} \nonumber \\
&& \hspace{-0.5cm} \times \int_0^{\infty} d k_{\perp} d k_{1\perp} \, 
\tW_{\Theta}(k_{\perp} r) \tW_{\Theta_1}(k_{1\perp} r) \tW_{\Theta_2}(k_{1\perp} r) 
\nonumber \\
&& \hspace{-0.5cm} \times k_{1\perp} P_{L0}(k_{\perp}) 
P_{g_1,e}(k_{1\perp},\eta) | \vec\theta-\vec\theta_2 | 
J_1(k_{\perp} r | \vec\theta-\vec\theta_2 |) \nonumber \\ 
&& \hspace{-0.5cm} \times J_0(k_{1\perp} r | \vec\theta_1-\vec\theta_2 |) .
\label{xi3-kSZ-perp}
\eeqa

It is useful to estimate the orders of magnitude of the three contributions
$\xi_{3\perp}, \xi_{3\parallel}^{\parallel}$ , and $\xi_{3\parallel}^{\perp}$.
Using $\tilde{W}'_{\Theta}(\ell) \sim \Theta \tilde{W}_{\Theta}(\ell)$,
and considering the case where we only have two angular scales
for the angles (\ref{conditions-real-space}),
\beq
\Theta_1 \sim \Theta_2 \sim | \vec\theta_1 - \vec\theta_2 | , \;\;\;
\Theta \sim | \vec\theta - \vec\theta_1 | \simeq | \vec\theta - \vec\theta_2 | ,\;\;\;
\Theta_1 \ll \Theta,
\label{angles-hierarchy}
\eeq
the transverse wavenumbers are of order
$k_{\perp} \sim 1/r\Theta$ and $k_{i\perp} \sim 1/r\Theta_i$.
This gives
\beq
\xi_{3\perp} \sim b_g I_g I_{g_1} I_{\rm kSZ_2} D^2 \Theta_2 k_{\perp} k_{2\perp}^2 
P_{L0}(k_{\perp}) P_{g_1,e}(k_{2\perp}) ,
\eeq
\beq
\xi_{3\parallel}^{\parallel} \sim b_g I_g I_{g_1} I_{\rm kSZ_2} \frac{D^2}{\eta} k_{1\perp}^2 
P_{L0}(k_{\perp}) P_{g_1,e}(k_{1\perp}) ,
\eeq
and
\beq
\xi_{3\parallel}^{\perp} \sim b_g I_g I_{g_1} I_{\rm kSZ_2} D^2 k_{\perp} k_{1\perp}^2 
| \vec\theta-\vec\theta_2 | P_{L0}(k_{\perp}) P_{g_1,e}(k_{1\perp}) ,
\eeq
hence
\beq
\frac{\xi_{3\perp}}{\xi_{3\parallel}^{\parallel}} \sim \Theta_2 k_{\perp} \eta 
\sim \frac{\Theta_2}{\Theta} \ll 1 , \;\;\;
\frac{\xi_{3\parallel}^{\perp}}{\xi_{3\parallel}^{\parallel}} 
\sim | \vec\theta - \vec\theta_2 | k_{\perp} \eta \sim 1 .
\label{xi3-kSZ-comparison}
\eeq
Thus, we find that the contribution $\xi_{3\perp}$ associated with the second term in
Eq.(\ref{kSZ-smooth}), which is due to the angle between the lines of sight within
the small conical beam of angle $\Theta_2$, is negligible as compared with the
contribution $\xi_{3\parallel}$ associated with the first term in
Eq.(\ref{kSZ-smooth}), which is the zeroth-order term.
However, the two components $\xi_{3\parallel}^{\parallel}$ and $\xi_{3\parallel}^{\perp}$
are of the same order.
The first one, $\xi_{3\parallel}^{\parallel}$, is the zeroth-order contribution when the lines of
sight $\vn$ and $\vn_2$ are taken to be parallel, whereas the second one, 
$\xi_{3\parallel}^{\perp}$, is the first-order contribution over this small angle,
measured by $| \vec\theta-\vec\theta_2|$ (which is, however, much larger than
the width $\Theta_2$ that gives rise to $\xi_{3\perp}$).
This first-order contribution can be of the same order as the zeroth-order contribution
because the latter is suppressed by the radial integration along the line of sight,
which damps longitudinal modes, $k_{\parallel} \ll k_{\perp}$.

In contrast with Eq.(\ref{xi3-J1}), the kSZ three-point correlation, given by the
sum of Eqs.(\ref{kSZ-transverse-1}), (\ref{xi3-kSZ-parallel}), and (\ref{xi3-kSZ-perp}),
does not vanish for orthogonal directions between the small-scale separation 
$(\vec\theta_1-\vec\theta_2)$ and the large-scale separation $(\vec\theta-\vec\theta_2)$.
Indeed, the leading order contribution in the squeezed limit to the response of
$\langle \delta_1 \vp_2 \rangle$ to a large-scale perturbation $\delta$ factors out as
$\langle \delta_1 \delta_2 \rangle \vv_{\delta}$, where we only take into account the contribution
that does not vanish at equal times (and we discard the finite-size smoothing effects). 
The intrinsic small-scale correlation 
$\langle \delta_1 \delta_2 \rangle$ does not depend on the large-scale
mode $\delta$, whereas $\vv_{\delta}$ is the almost uniform velocity due to the large-scale
mode, which only depends on the direction to $\delta(\vec\theta)$ and is independent of the
orientation of the small-scale mode $(\vec\theta_1-\vec\theta_2)$.

Because the measurement of the kSZ effect only probes the radial velocity of the
free electrons gas along the line of sight, which is generated by density fluctuations
almost parallel to the line of sight over which we integrate and which are damped by this
radial integration, the result (\ref{xi3-kSZ-parallel})
is suppressed as compared with the ISW result (\ref{xi3-J1})
by the radial derivative $d\ln(b_g I_g D)/d\eta \sim 1/r$.
Also, the contribution (\ref{xi3-kSZ-parallel}), associated with 
transverse fluctuations that are almost orthogonal to the second line of sight,
is suppressed as compared with the ISW result (\ref{xi3-J1}) by the 
small angle $| \vec\theta-\vec\theta_2 |$ between the two lines of sight.

One drawback of the kSZ consistency relation, (\ref{kSZ-transverse-1}) and
(\ref{xi3-kSZ-parallel})-(\ref{xi3-kSZ-perp}),
is that it is not easy to independently measure the galaxy-free electrons power spectrum 
$P_{g_1,e}$, which is needed if we wish to test this relation.
Alternatively, Eqs.(\ref{xi3-kSZ-parallel})-(\ref{xi3-kSZ-perp})
may be used as a test of models for the free electrons
distribution and the cross power spectrum $P_{g_1,e}$.

\subsection{Lensing-lensing-kSZ correlation}
\label{sec:Lensing-lensing-kSZ}

Again, from Eqs.(\ref{kSZ-transverse-1}) and (\ref{xi3-kSZ-parallel})-(\ref{xi3-kSZ-perp})
we can directly obtain the lensing-lensing-kSZ three-point correlation,
\beq
\xi_3(\kappa^s,\kappa^s_1,\Delta^s_{\rm kSZ_2}) = \langle \kappa^s(\vec\theta) \,
\kappa^s_1(\vec\theta_1) \, \Delta^s_{\rm kSZ_2}(\vec\theta_2) \rangle ,
\label{xi3-lens-lens-kSZ}
\eeq
by replacing the galaxy kernels $b_g I_g$ and $I_{g_1}$ by the lensing convergence 
kernels $I_{\kappa}$ and $I_{\kappa_1}$.
This gives 
$\xi_3 = \xi_{3\perp} + \xi_{3\parallel}^{\parallel} + \xi_{3\parallel}^{\perp}$ with
\beqa
\xi_{3\perp} & = & \frac{(\vec\theta-\vec\theta_1) \cdot (\vec\theta_2 - \vec\theta_1)}
{| \vec\theta -\vec\theta_1 |  | \vec\theta_2 - \vec\theta_1 |} (2\pi)^4
\int d\eta \, I_{\kappa} I_{\kappa_1} I_{\rm kSZ_2} \frac{d\ln D}{d\eta} \nonumber \\
&& \hspace{-0.5cm} \times \int_0^{\infty} d k_{\perp} d k_{2\perp} \, k_{2\perp}
\tW_{\Theta}(k_{\perp} r) \tW_{\Theta_1}(k_{2\perp} r) \tW_{\Theta_2}'(k_{2\perp} r) 
\nonumber \\
&& \hspace{-0.5cm} \times P_L(k_{\perp},\eta) P_{m,e}(k_{2\perp},\eta)
J_1(k_{\perp} r | \vec\theta-\vec\theta_1 |) \nonumber \\ 
&& \hspace{-0.5cm} \times J_1(k_{2\perp} r | \vec\theta_2-\vec\theta_1 |) ,
\label{kSZ-transverse-1-lens-lens}
\eeqa
\beqa
\xi_{3\parallel}^{\parallel} & = & - (2\pi)^4 \int d\eta \, \frac{d}{d\eta} \left[ I_{\kappa} D \right] 
I_{\kappa_1} I_{\rm kSZ_2} \frac{dD}{d\eta} \int_0^{\infty} d k_{\perp} d k_{1\perp} 
\nonumber \\
&& \hspace{-0.6cm} \times \tW_{\Theta}(k_{\perp} r) \tW_{\Theta_1}(k_{1\perp} r) 
\tW_{\Theta_2}(k_{1\perp} r) \frac{k_{1\perp}}{k_{\perp}} P_{L0}(k_{\perp}) 
\nonumber \\
&& \hspace{-0.6cm} \times P_{m,e}(k_{1\perp},\eta) J_0(k_{\perp} r | \vec\theta-\vec\theta_2 |)  
J_0(k_{1\perp} r | \vec\theta_1-\vec\theta_2 |) ,
\label{xi3-kSZ-parallel-lens-lens}
\eeqa
and
\beqa
\xi_{3\parallel}^{\perp} & = & - (2\pi)^4 \int d\eta \, I_{\kappa} I_{\kappa_1} I_{\rm kSZ_2} D 
\frac{dD}{d\eta} \int_0^{\infty} d k_{\perp} d k_{1\perp} \nonumber \\
&& \hspace{-0.9cm} \times \tW_{\Theta}(k_{\perp} r) 
\tW_{\Theta_1}(k_{1\perp} r) \tW_{\Theta_2}(k_{1\perp} r) k_{1\perp} P_{L0}(k_{\perp}) 
\nonumber \\
&& \hspace{-0.9cm} \times P_{m,e}(k_{1\perp},\eta) | \vec\theta-\vec\theta_2 | 
J_1(k_{\perp} r | \vec\theta-\vec\theta_2 |) J_0(k_{1\perp} r | \vec\theta_1-\vec\theta_2 |) 
. \;\;\; \nonumber \\
&&
\label{xi3-kSZ-perp-lens-lens}
\eeqa
This now involves the matter-free electrons cross power spectrum $P_{m,e}$.

The application of the relations above is, unfortunately, a nontrivial task in terms of observations: 
to test those relations one would require the mixed galaxy (matter) - free electrons power spectrum. 
One possibility would be to do a stacking analysis of several X-ray observations of the hot ionized 
gas by measuring the bremsstrahlung effect. 
For instance, one could infer $n_e n_p T^{-1/2}$, by making some reasonable assumptions 
about the plasma state, as performed in \citet{fil}, with the aim of measuring $n_e$ in filaments.
We would of course need to cover a large range of scales. For kpc scales, inside galaxies 
and in the intergalactic medium, one could use for instance silicon emission line ratios 
\citep{neb1,neb2}. 
For Mpc scales, or clusters, one may use the Sunyaev-Zeldovich (SZ) effect \citep{planck}. 
Nevertheless, all these proposed approaches are quite speculative at this stage.

\subsection{Suppressed contribution to the galaxy-kSZ-kSZ correlation}
\label{sec:Galaxy-kSZ-kSZ}

As for the ISW effect, we investigate whether the galaxy-kSZ-kSZ correlation provides
a good probe of the consistency relations.
For the same symmetry reasons as in Section~\ref{sec:Galaxy-ISW-ISW}, 
we find that the leading-order contribution to this three-point correlation vanishes.
Let us briefly sketch how this cancellation appears.
First, from the hierarchy (\ref{xi3-kSZ-comparison}) we neglect the contribution
associated with the second term in Eq.(\ref{kSZ-smooth}), that is, the widths of the small-scale
windows are small and we can approximate each conical beam as a cylinder
(flat-sky limit).
Then, we only have the component $\xi_{3\parallel \; \parallel}$ similar to 
Eq.(\ref{kSZ-consistency-parallel-0}), which gives in compact notations
\beq
\langle \delta \Delta_{\rm kSZ_1} \Delta_{\rm kSZ_2} \rangle \propto \langle
\tilde\delta(\vk) [\vn_1\cdot\tilde\vp_e(\vk_1)] [\vn_2\cdot\tilde\vp_e(\vk_2)] \rangle' .
\eeq
The consistency relation (\ref{consistency_relation_p}) gives at equal times
\beqa
\langle \delta \Delta_{\rm kSZ_1} \Delta_{\rm kSZ_2} \rangle & \propto &
\frac{\vn_1\cdot\vk}{k^2} \langle \tilde\delta_e(\vk_1) [\vn_2\cdot\tilde\vp_e(\vk_2)] 
\rangle'  \nonumber \\
&& + \frac{\vn_2\cdot\vk}{k^2} \langle [\vn_1\cdot\tilde\vp_e(\vk_1)] \tilde\delta_e(\vk_2) 
\rangle' .
\eeqa
In the regime (\ref{angles-hierarchy}), we can take $\vn_1 \simeq \vn_2$,
hence
\beqa
\langle \delta \Delta_{\rm kSZ_1} \Delta_{\rm kSZ_2} \rangle \!\!\!\! & \propto & \!\!\!\!
\frac{\vn_1\!\cdot\!\vk}{k^2} \vn_1 \!\cdot\! \left[ \langle \tilde\delta_e(\vk_1) \tilde\vp_e(-\vk_1) 
\rangle' + \langle \tilde\vp_e(\vk_1) \tilde\delta_e(-\vk_1) \rangle' \right] \nonumber \\
& = & 0 .
\eeqa
Here we used the fact that the density-momentum cross power spectrum
obeys the symmetry 
$\langle \tilde\delta_e(\vk) \tilde\vp_e(-\vk) \rangle' = - 
\langle \tilde\delta_e(-\vk) \tilde\vp_e(\vk) \rangle'$, associated with a change of sign of the 
coordinate axis.

This cancellation can again be understood in configuration space. 
At leading order in the squeezed limit, the linear change of 
$\langle p_{\parallel}(\vec\theta_1) p_{\parallel}(\vec\theta_2) \rangle'$
due to a large-scale perturbation $\delta(\vec\theta)$ is
$(\langle \delta_1 p_{\parallel 2} \rangle' + \langle p_{\parallel 1} \delta_2 \rangle') 
v_{\delta\parallel}$, where $\vv_{\delta}$ is the large-scale velocity generated by the
large-scale mode (the second-order term 
$\langle 1+\delta_1 \delta_2 \rangle v_{\delta\parallel}^2$ does not contribute to the
response function and the consistency relation).
By symmetry the sum in the parenthesis vanishes.
Therefore, in this paper we focus on the three-point correlations
(\ref{correlation-delta-line-kSZ}) and (\ref{xi3-lens-lens-kSZ}), with only one kSZ field.

\section{Conclusions}
\label{sec:Conclusions}

In this paper, we have shown how to relate the large-scale consistency relations with
observational probes. Assuming the standard cosmological model (more specifically,
the equivalence principle and Gaussian initial conditions), nonzero equal-times consistency
relations involve the cross-correlations between galaxy or matter density fields
with the velocity, momentum, or time-derivative density fields. We have shown that
these relations can be related to actual measurements by considering the ISW and kSZ
effects, which indeed involve the time derivative of the matter density field and the
free electrons momentum field.
We focused on the lowest-order relations, which apply to three-point correlation functions 
or bispectra, because higher-order correlations are increasingly difficult to measure.

The most practical relation obtained in this paper is probably the one associated with the
ISW effect, more particularly its cross-correlation with two cosmic weak-lensing convergence
statistics.
Indeed, it allows one to write this three-point correlation function in terms of two 
matter density field power spectra (linear and nonlinear), which can be directly measured 
(e.g., by two-point weak lensing statistics). 
Moreover, the result, which is the leading-order contribution in the squeezed limit,
shows a specific angular dependence as a function of the relative angular positions 
of the three smoothed observed statistics.
Then, both the angular dependence and the quantitative prediction provide 
a test of the consistency relation, that is, of the equivalence principle and of primordial
Gaussianity.
If we consider instead the cross-correlation of the ISW effect with two galaxy density 
fields, we obtain a similar relation but it now involves the mixed galaxy-matter density 
power spectrum $P_{g,m}$ and the large-scale galaxy bias $b_g$. These two quantities 
can again be measured (e.g., by two-point galaxy-weak lensing statistics) and provide 
another test of the consistency relation.

The relations obtained with the kSZ effect are more intricate. They do not show a simple
angular dependence, which would provide a simple signature, and they involve the 
galaxy-free electrons or matter-free electrons power spectra. These power spectra are
more difficult to measure. One can estimate the free electron density in specific regions,
such as filaments or clusters, through X-ray or SZ observations, or around typical 
structures by stacking analysis of clusters. 
This could provide an estimate of the free electrons cross power spectra and a check 
of the consistency relations. Although we can expect significant error bars, it would be 
interesting to check that the results remain consistent with the theoretical predictions.
A violation of these consistency relations would signal either a modification of gravity
on cosmological scales or non-Gaussian initial conditions. We leave to future works
the derivation of the deviations associated with various nonstandard scenarios.

\begin{acknowledgements}

This work is supported in part by the French Agence Nationale de la Recherche
under Grant ANR-12-BS05-0002. DFM thanks the support of the Research Council of Norway.

\end{acknowledgements}

\bibliographystyle{aa} 
\bibliography{ref1}   

\end{document}